\documentclass[useAMS,usenatbib]{mn2e}
\usepackage{graphicx}
\title[MASH-II]{MASH-II: More Planetary Nebulae from the AAO/UKST H$\alpha$ Survey}
\author[B. Miszalski et al.]{Brent Miszalski$^{1,2}$\thanks{E-mail: brent@ics.mq.edu.au}, Q. A. Parker$^{1,3}$, A. Acker$^{2}$, J. L. Birkby$^{4}$, D. J. Frew$^{1,5}$ \newauthor and A. Kovacevic$^{1}$\\ 
$^{1}$Department of Physics, Macquarie University, Sydney, NSW 2109, Australia\\
$^{2}$Observatoire Astronomique, Universit\'e Louis Pasteur, 67000 Strasbourg, France\\
$^{3}$Anglo-Australian Observatory, Epping, NSW 1710, Australia\\
$^{4}$Institute of Astronomy, University of Cambridge, Madingley Road, Cambridge CB3 OHA\\
$^{5}$Perth Observatory, Bickley, WA 6076, Australia\\
}
\begin{document}

\date{Accepted 2007 November 15.  Received 2007 November 15; in original form 2007 October 9}

\maketitle

\begin{abstract}
We present a supplement to the Macquarie/AAO/Strasbourg H$\alpha$ planetary nebulae (PNe) catalogue (MASH), which we denote MASH-II. 
The supplement consists of over 300 true, likely and possible new Galactic PNe found after re-examination of the entire AAO/UKST H$\alpha$ survey of the southern Galactic Plane in digital form.
We have spectroscopically confirmed over 240 of these new candidates as bona-fide PNe and we include other high quality candidates awaiting spectroscopic confirmation as possible PNe.
These latest discoveries largely comprise two distinct groups: small, star-like or moderately resolved PNe at one end and mostly large, extremely low surface brightness PNe at the other.
Neither group were easy to discover from simple visual scrutiny of the original survey exposures as for MASH but were relatively straightforward to uncover from the digital images via application of semi-automated discovery techniques. 
We suspect the few PNe still hidden in the H$\alpha$ survey will lie outside our search criteria or be difficult to find.
\end{abstract}

\begin{keywords}
astronomical data bases: miscellaneous - catalogues - surveys - planetary nebulae: general 
\end{keywords}
\section{Introduction}
Planetary Nebulae (PNe) are a very brief ($\sim$10,000--50,000 year) phenomenon exhibited by most stars of low 
to intermediate mass after their post AGB phase when the previously ejected shroud of enriched and processed stellar material 
becomes ionised by the hot, remnant stellar core as it evolves towards a white dwarf.
PNe are an essential means of understanding the late evolution of 
low-mass stars like our Sun.
PNe hold the key to determining the physics of mass loss and time-scales of low
mass stellar evolution. 
Since low mass stars can each recycle a significant fraction of their mass into the ISM, PNe are vital probes of stellar nucleosynthesis processes and ISM enrichment. Their strong emission line spectra make them visible at large distances, enabling kinematics and abundances of elements generally difficult to measure in stars to be determined. 

The properties of their progenitor stars can be investigated and can yield important information on core mass, temperature and stellar evolutionary history which ties into their observed nebula properties. 
If distances can be determined for example, from measured parallaxes (Harris et al. 2007) or photometric parallaxes (Ciardullo et al. 1999), by assuming a fixed distance in other galaxies (e.g. Reid \& Parker 2006), or more generally via statistical methods such as the new H$\alpha$ surface brightness radius relation of Frew \& Parker (2006), then observed properties can unlock true physical parameters. These include the size of the inner shell and outer faint AGB halo (Corradi et al. 2003), total nebular mass and central star brightness and evolutionary state.

The PNe population of our Galaxy from observation based estimates is 23,000$\pm$6,000 disc PNe (Zijlstra \& Pottasch 1991) or 28,000$\pm$5,000 in total (Frew \& Parker 2006). A larger value of 46,000$\pm$13,000 PNe has been obtained from stellar population synthesis models (Moe \& De Marco 2006), though this is only for PNe with radii $<$ 0.9 pc and so is not strictly comparable with the aforementioned estimates. It is however clear that all estimates far exceed the number of PNe actually catalogued so far ($\sim$2,500) and that Galactic numbers are dwarfed by large numbers of known extragalactic PNe (Magrini 2006). Such disparity can be attributed to a combination of heavy obscuration by dust in the Galactic Plane, the short lifetime of PNe, and the extraordinary diversity in observed PN properties.
The observables which exhibit the most diversity, and thus influence any survey aiming for high completeness the most, are the surface brightness, angular extent and morphology (that may have been modified by an interaction with the ISM).
To address the discrepancy in PN numbers sky surveys with significant areal coverage, sensitivity and angular resolution over suitable passbands are thus required to uniformly sample the whole population.

Historically, surveys have had difficulty meeting these requirements simultaneously and were often restricted to one wavelength domain (for an extensive overview see Parker et al. 2006a). This has led to significant biases and incompleteness in PNe catalogues limiting the range of observed evolutionary diversity. Substantial progress has been made recently in addressing these biases with the advent of modern H$\alpha$ surveys such as the SuperCOSMOS H$\alpha$ survey (see Section \ref{sec:mash}) and the INT Photometric H$\alpha$ Survey of the Northern Galactic Plane (IPHAS; Drew et al. 2005). A search for PNe within IPHAS is well underway (e.g. Mampaso et al. 2006, Viironen et al. 2006).
For these reasons we expect large numbers to have eluded detection in previous surveys, and coupled with the aforementioned scientific drivers, we are strongly motivated to search for undiscovered PNe in our Galaxy.

This paper is structured as follows. Section \ref{sec:mash} introduces the MASH PNe catalogue focusing on previous discovery techniques. Section \ref{sec:mashii} describes our candidate selection and visualisation techniques. Section \ref{sec:spec} discusses the spectroscopic confirmation of candidates which leads to the presentation of the MASH-II PNe catalogue in Section \ref{sec:cat}. Completeness is discussed in Section \ref{sec:comp} and we conclude in Section \ref{sec:conc}.

\section{The MASH Planetary Nebulae Catalogue: Discovery Techniques}
\label{sec:mash}
The recent publication of the Macquarie/AAO/Strasbourg H$\alpha$ PNe catalogue 
(MASH; Parker et al. 2006b, hereafter Paper~I) presented $\sim$900 new, spectroscopically confirmed Galactic PNe. MASH has boosted the number of Galactic PNe by nearly 60 per cent and offers considerable scope to address problems in PNe research afresh.

The majority of the significant MASH PNe were discovered from careful visual examination of the Anglo-Australian/UK Schmidt Telescope SuperCOSMOS H$\alpha$ survey of 
the Southern Galactic Plane (SHS\footnote{http://www-wfau.roe.ac.uk/sss/halpha}; Parker et al. 2005). All 233 SHS fields were viewed as original film media under a microscope or eyepiece.
This was effective at finding resolved PNe based on their H$\alpha+$[NII] nebular morphology, which is often considerably weaker or absent in the matching short-red (SR) broad-band red exposure. 
This painstaking work was later complemented by selective analysis of the digital SHS pixel data when it became available in 2003. Initially it was used to revise candidate position, size and morphology, however it was later employed to conduct two searches for additional PNe missed by the visual search. 
The first digital search looked for small, star-like PNe in 18 Galactic Bulge fields, whilst the second focused on large, faint PNe using 16 times blocked-down images of whole survey fields (see Paper~I for further details).

However the additional searches had their limitations. 
The search for small PNe predominantly used PSF matched difference images (Bond et al. 2001) and only 18 Galactic Bulge fields were searched. The search for large PNe involved visual inspection of the digital blocked-down H$\alpha$ survey fields (11 arcsec resolution) whilst altering the contrast to enhance low surface brightness nebulosity. Many PNe candidates were recovered that were too large for the microscope field-of-view during the original film inspection, but high stellar density in many fields limited its sensitivity.
\section{MASH-II}
\label{sec:mashii}
Here we report on further PNe discoveries made after Paper~I, which we denote as MASH-II.
MASH-II differs from MASH in its semi-automated discovery techniques that were applied to the on-line SHS digital data of all 233 fields for the first time. MASH-II consists of two different stages. 
The first and most fruitful stage, in terms of the total number of candidates found, used the Image Analysis Mode (IAM) photometric data to target star-like or compact PNe (Section \ref{sec:iam}). 
The second stage used blocked-down \emph{quotient} images of SHS fields to target low surface brightness PNe (Section \ref{sec:quotient}).
A key feature and strength of MASH-II was the incorporation of multi-wavelength data within these stages as part of the initial candidate selection process (Section \ref{sec:vis}). Results from the searches are discussed in Section \ref{sec:results}.

\subsection{Candidate Selection}
\label{sec:selection}
\subsubsection{IAM photometry}
\label{sec:iam}
Early on it was realised that the IAM photometric data generated by the SuperCOSMOS pipeline (Hambly et al. 2001) would provide an excellent avenue for PN discovery (e.g. Pierce 2005).
However, prior to MASH-II, the IAM data has not been used extensively to discover new PNe.
Chance discoveries of three PNe were made during colour-cut searches for emission line stars by Pierce (2005) and R. Pretorius (see Paper~I). 
G\'orny (2006) used IAM photometry to discover 24 PNe, nine of which were new, located in the direction of the Galactic Bulge.

\begin{figure}
   \begin{center}
      \includegraphics[scale=1.0]{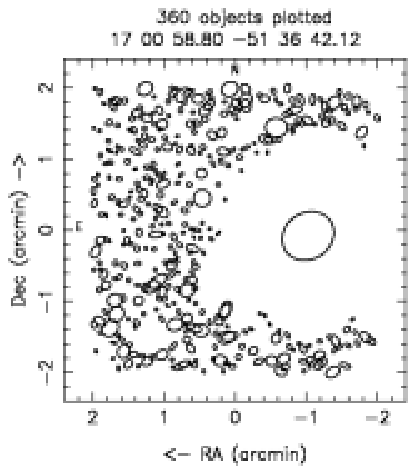}
      \includegraphics[scale=1.0]{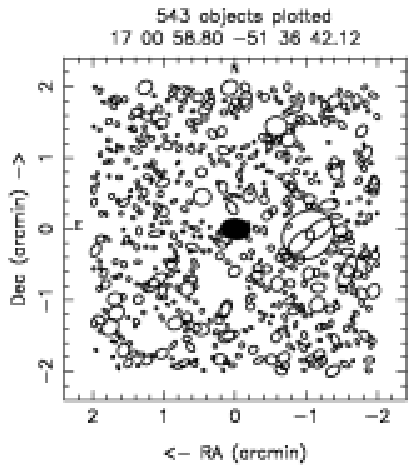}
   \end{center}
   \caption{The effect of varying the image quality threshold parameter for the survey IAM data. 
The default value of 127 (left) is clearly unsatisfactory near bright stars, however by selecting 2048 (right) 
the new MASH-II PN MPA1700-5136 is recovered (filled ellipse).}
   \label{fig:catquality}
\end{figure}

The IAM photometric data are available as a FITS table extension in all SHS pixel data, however for heavy-duty applications the data are available online as a separate data product\footnote{http://www-wfau.roe.ac.uk/sss/halpha/haobj.html} which we use here. For this project the central coordinates of all 233 survey fields were used to extract large photometric catalogues in the H$\alpha$ waveband of either $240 \times 240$ arcmin 
(central survey fields) or $360 \times 360$ arcmin (peripheral survey fields). Following Pierce (2005), the `expert' catalogue extraction parameter `image quality threshold' was set to 2048 which includes deblended objects affected by bright stars (see Hambly et al. 2001). This offers significant benefits near bright stars (Fig. \ref{fig:catquality}), however it also increases the number of spurious deblends. In our case this was not a problem except in large areas of diffuse emission such as HII regions where large numbers of spurious candidates had to be processed which slowed down our search in some fields. 
Nevertheless, we persisted with our deblending prescription as many serendipitous discoveries of very faint PNe were made following deblending of their bright rims.
We did not change any of the other catalogue extraction parameters from their default values.

Rather than calculating an optimal cut in H$\alpha-$SR for each field to find H$\alpha$ emitters likely to be PNe, we opted for a more generic approach consisting of several cuts applied to all fields equally. 
We tabulate the candidate selection criteria applied to the IAM photometry for each cut in Table \ref{tab:cuts}. Our first cut was designed to find the bulk of our expected PNe population as the majority of MASH PNe are brighter than 16th mag in H$\alpha$. The limit was a compromise between finding sufficient numbers of PNe and reducing the numbers of SHS film artefacts included in the cuts. On the order of a few hundreds of candidates were visually examined per field in the first cut using the method described in Section \ref{sec:vis}. This cut resulted in the bulk of our new candidate PNe (Table \ref{tab:cuts}).

\begin{table}
   \begin{center}
  \begin{tabular}{llllll}
      \hline
      Cut & H$\alpha-$SR & H$\alpha$ & ellipticity & area & candidates \\
      \hline
      1    & $\leq$ -2 & $\leq$ 16 & - & - & 300\\
      2    & (-2,-1.5] & $\leq$ 16 & $\leq$ 0.25 & $\geq$ 100 & 50\\
      3    & (-1.5,-1.25] & - & $\leq$ 0.25 & $\geq$ 100 & 20\\
      4$^a$ & SR undefined & $\leq$ 17 & $\leq$ 0.8 & $\geq$ 50 & 10--20 \\
      \hline
   \end{tabular} 
\end{center}
$^a$Quality flag 1 or 2 (star or galaxy).\\
\caption{Candidate selection criteria applied to IAM photometry for each of the 233 SHS fields. An estimate of the respective number of new candidate PNe found is given in the last column.}
   \label{tab:cuts}
\end{table}

To see if we could improve on the first cut, we made two further cuts aimed towards objects with less H$\alpha$ excess. However, the number of candidates to inspect per field grew substantially, up to two thousand per field, such that further constraints had to be applied. By constraining the ellipticity and area of deblended catalogue entries (see the respective columns in Table \ref{tab:cuts} and Hambly et al. 2001), we targeted rounder and more star-like PNe with real image sizes as the first cut was deemed to have picked up most of the small, resolved PNe.
These cuts did contribute some additional candidates, but as many fields were `empty' especially at high latitudes, diminishing returns influenced us to conduct no more cuts in H$\alpha-$SR.

\begin{figure}
   \begin{center}
      \includegraphics[scale=0.5]{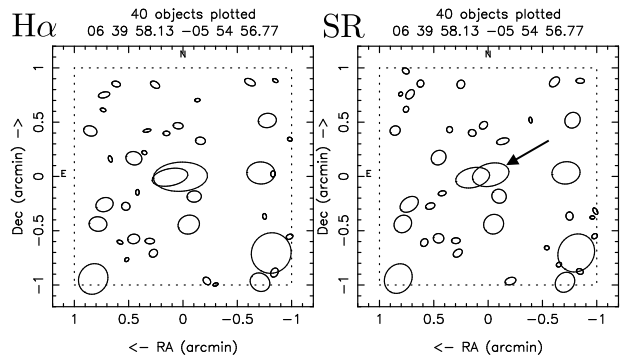}
   \end{center}
   \caption{The deblending around the new MASH-II PN MPA0639-0554 (arrowed) in the H$\alpha$ (left) and SR (right) wavebands. Each waveband is deblended separately which can lead to H$\alpha$ not being assigned a SR magnitude (and vice versa) when the deblending result is different in each waveband. In such cases the PN could not have been found using a simple H$\alpha-$SR cut.}
   \label{fig:nosr}
\end{figure}

The final cut in Table \ref{tab:cuts} was prompted after finding that three of the nine new PNe from G\'orny (2006) were not recovered in our initial cuts because they had no SR magnitude.
This can happen when either a faint PN has no real SR detection, or less commonly where crowding and the separate deblending in both H$\alpha$ and SR wavebands can cause the SR magnitude to be undefined in the H$\alpha$ waveband catalogue (Figure \ref{fig:nosr}).
This onerous final cut is ongoing and has only been applied to 78 per cent of the 233 SHS fields. Yields from the cut have been very small and hence we expect only very few additional PNe to be found upon completion of the cut.

\subsubsection{Blocked-down quotient images}
\label{sec:quotient}
\begin{figure}
   \begin{center}
      \includegraphics[scale=1.0]{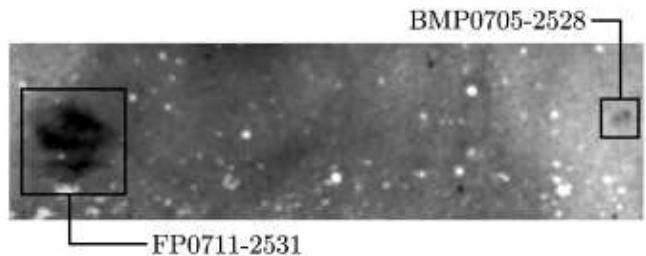}
   \end{center}
   \caption{A smoothed 90 $\times$ 25 arcmin$^2$ section of the blocked-down quotient image of SHS field HA756. The MASH PN FP0711-2531 was found during inspection of the blocked-down H$\alpha$ image of the field, whereas using the superior quotient image we have uncovered the much fainter new MASH-II PN BMP0705-2528 nearby. NE is to the top left-hand corner.}
   \label{fig:quotient}
\end{figure}

The blocked-down H$\alpha$ images of each SHS field (11 arcsec/pix) have already enabled the discovery of a number of very large, low surface brightness PNe (see Paper~I for details).
However, this work did not benefit from the availability of the matching blocked-down SR image.
We have since made use of these images to construct blocked-down \emph{quotient} images of each field (hereafter the `quotient image'), the division of H$\alpha$ by SR after image alignment. These quotient images are \emph{significantly} more powerful in revealing extended low surface brightness PNe than the H$\alpha$ image alone. In particular, the reduced stellar crowding has enabled many new, intermediate sized PNe (1--5 arcmin across) of extremely low surface brightness to be found (e.g. Figure \ref{fig:quotient}). 

A systematic search of all 233 fields was carried out by one of us (JB). The search focused on finding all large scale coherent nebulae as well as irregular and asymmetric nebulae. The \textsc{FITS} viewer \textsc{ds9}\footnote{http://hea-www.harvard.edu/RD/ds9} was used to display the quotient images making use of the many scaling options to sufficiently explore the wide dynamic range of these superb images. Software was developed to interface with \textsc{ds9} to allow `point and click' retrieval of full-resolution images in a pseudo real-time fashion with images displayed in a new window. Region files were used in \textsc{ds9} to indicate the positions of known PNe and to keep track of positions already queried for in SIMBAD.

\subsection{Candidate Management and Visualisation}
\label{sec:vis}
Following candidate selection a number of steps were taken to produce a preliminary, working list of new candidate PNe. All candidates from both searches were assigned a working name based on how they were discovered and accurate equatorial coordinates. Those found from the IAM photometry were assigned the prefix MPA (denoting the authors Miszalski, Parker and Acker), whilst those from the quotient images were assigned BMP (denoting the authors Birkby, Miszalski and Parker) or MPA. There are some exceptions however, namely some large MPA PNe, discovered by BM inspecting either full-resolution data or quotient images, and some small BMP PNe, visible as small groups of white pixels in the quotient images. Each prefix is completed with concatenated equatorial coordinates as for the PHR entries in the MASH catalogue of Paper~I. 

Before images could be retrieved, candidate lists from the IAM search had duplicate entries removed by using our adopted nomenclature as a unique identifier. 
This was generally acceptable as our images were chosen to be $2 \times 2$ arcmin$^2$ and the vast majority of candidates were centred in these images. 
In the quotient image search the size of the images were chosen to suit the size of each candidate.

Use of the current SHS batch retrieval form to download full-resolution H$\alpha$ and SR pixel data proved too onerous for general use because of the large numbers of candidates to be processed.
To overcome this a special script was developed to read in the candidate lists 
and submit the coordinates to the SHS image retrieval form one at a time. 
The script also proceeded to download the UKST SuperCOSMOS B$_\mathrm{J}$ image, the 2MASS J, H and K images, and once downloads were complete, to construct not only the quotient H$\alpha$/SR image, but also colour images which enhance our PNe detection capability (described below). A webpage was generated by the script from the candidate coordinate lists, containing the three image previews alongside links to SIMBAD and a local database query to check for already known PNe.

Much of our enhanced PNe detection capability can be attributed to the B$_\mathrm{J}$ image.
The B$_\mathrm{J}$ image has a plate limit of B$_J\sim$23 (Hambly et al. 2001) that allows for the routine detection of blue central star (CS) candidates as faint as mag 20.
A blue CS candidate in an otherwise inconspicuous region of extended H$\alpha$ emission can add significant weight to the veracity of a candidate PN, especially if found near the geometric centre, that might have otherwise been discounted as an HII region. Stromgren spheres may also be found around hot white dwarfs or subdwarfs (i.e. faint blue stars), though these can often be discounted based on very irregular or atypical morphologies (see Frew \& Parker 2006, Madsen et al. 2006). The B$_\mathrm{J}$ image can also act as a shallow [OIII] and H$\beta$ image as these lines are typically strong in PNe, unless heavily reddened. 

\begin{figure*}
   \begin{center}
         \includegraphics[scale=0.95]{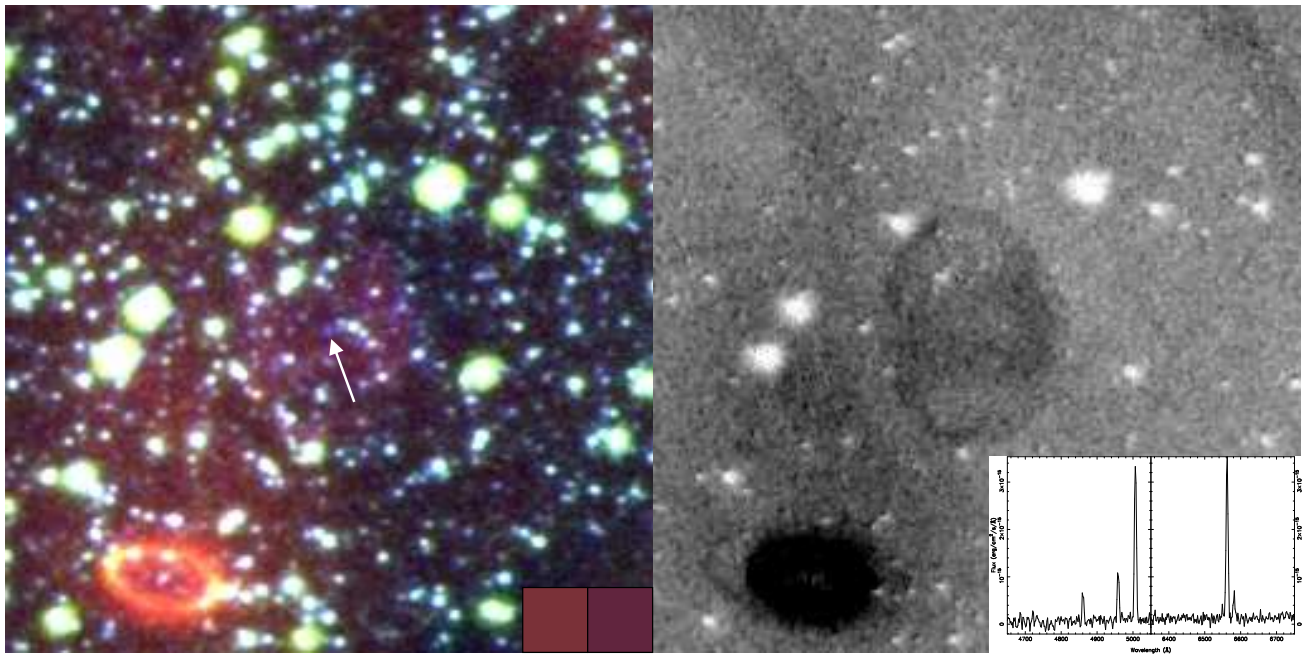}
   \end{center}
\caption{The new MASH-II PN MPA1644-4002 (PN G343.6+03.7a) depicted in the colour composite (left) and quotient image (right). 
   Note the distinctive purple hue and sharp eastern edge that separates the new PN from the ruddy coloured background emission (clarified by small squares). 
   A B$_J$=20.8 CS candidate (arrowed) and faint extensions (right) can also be seen.
   The spectrum exhibits relatively strong [OIII] (lower right).
   Its low surface brightness has prevented its discovery until now, despite being only 1.3 arcmin north west of known PN G343.6+03.7 (SuWt 3; West \& Schuster, 1980). NE is to the top left-hand corner.}

   \label{fig:colour}
\end{figure*}

These strengths of the B$_\mathrm{J}$ image are best realised by creating false-colour 
composite images with red, green and blue channels taken from H$\alpha$, SR and B$_\mathrm{J}$ 
images respectively (hereafter the `colour composite'). 
Such images can be created using either the IRAF task \textsc{export} or more interactively with the RGB frame capability of \textsc{ds9}. 
In such images PNe generally stand out clearly against a field of comparatively neutral stars with a pink or purple hue, due to contributions from both H$\alpha$ and [OIII], or with an orange or red hue due to weaker/absent [OIII] emission or high reddening.
Such colours can help separate diffuse PNe from diffuse background H$\alpha$ emission which is often a ruddy colour (Figure \ref{fig:colour}) and can also act as a rough proxy for excitation class where the extinction is low. A similar technique has been used to examine the mid infra-red properties of MASH PNe detected in the Spitzer/GLIMPSE survey (Cohen et al. 2007).

As an adjunct to the colour composite we also used an equivalent 2MASS image with the
red, green and blue channels now formed from the 2MASS K$_s$, H and J atlas images 
respectively (hereafter the `2MASS image'). The 2MASS image helped in eliminating at a glance unusually strong (K$_s \la 10$) NIR point-source emitters such as emission line stars, symbiotic stars and compact HII regions which can appear very similar to compact PNe in the optical. Further refinement in the NIR was left till later when spectra were available to de-redden 2MASS magnitudes and to assess other criteria (see Paper~I). The 2MASS image can also portray a faint CS, resolved NIR emission or no emission for PNe.

\subsection{Results}
\label{sec:results}
A total of $\sim$550 new candidates were found in the first instance, comprised of $\sim$400 PNe from the IAM search (70 per cent) and $\sim$150 PNe from the blocked-down quotient image search ($\sim$30 per cent).
The latter sample is made up of equal parts small, high surface brightness nebulae and large, low surface brightness nebulae.
The candidates were incrementally added to a local database upon various initial criteria such as an obvious H$\alpha$ excess (as described in Section \ref{sec:selection}), apparent PN morphology, blue central star candidate, and 2MASS K$_s > 10$.
New additions to the database were checked against our local database and SIMBAD, although some (but not all) PNe listed as `possible' in SIMBAD lacking confirmatory spectroscopy were also added. 
Our sample was later refined to $\sim$350 candidates prior to confirmatory spectroscopy based on a more detailed analysis of the accrued sample following the criteria in Paper~I. 

\section{Spectroscopic Confirmation of MASH-II Candidates}
\label{sec:spec}
\begin{figure*}
   \begin{center}
      \includegraphics[scale=1.0]{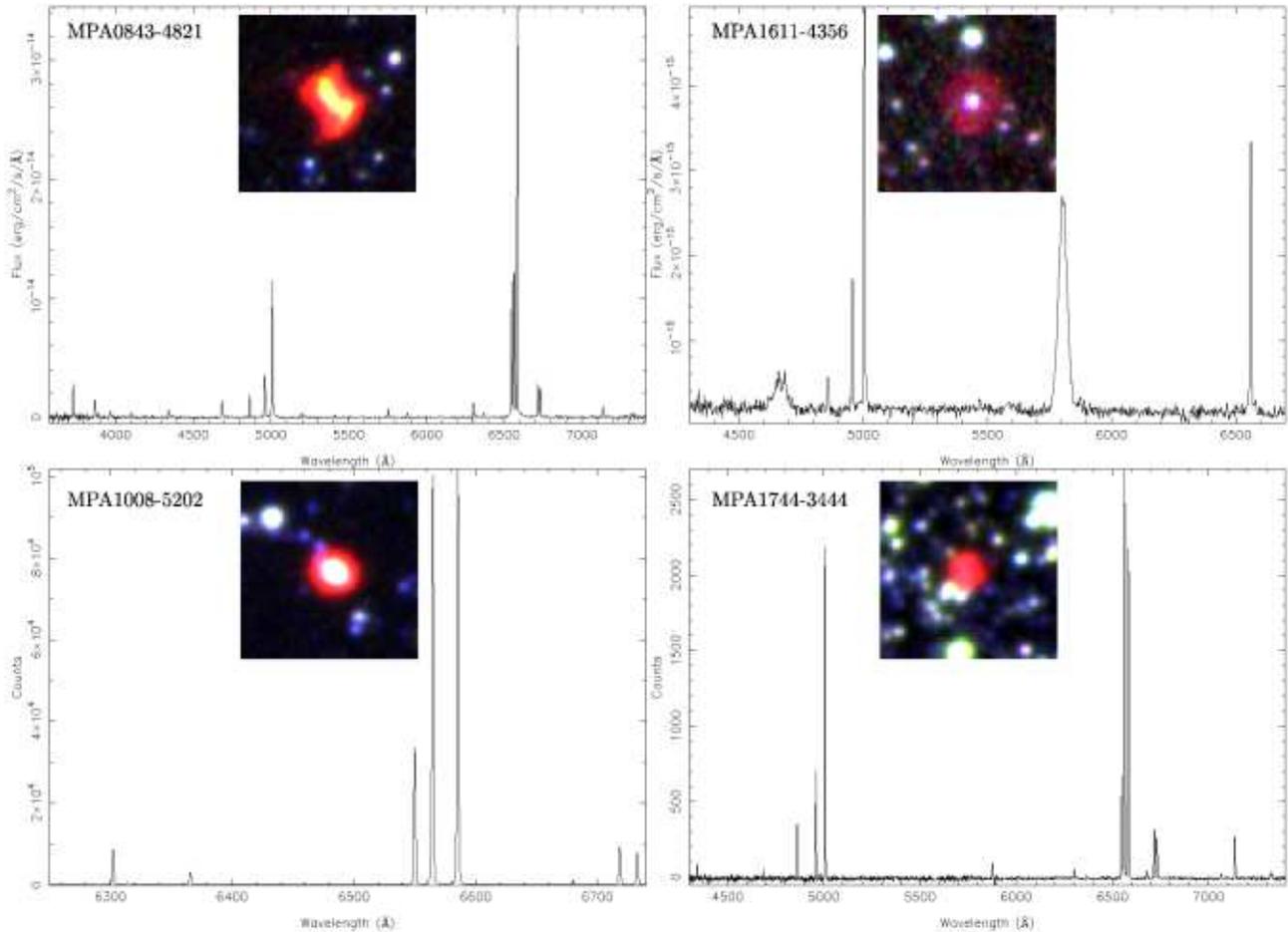}
   \end{center}
   \caption{An assortment of MASH-II PNe spectra and colour composite images. Included are spectra from the ANU MSSSO 2.3-m telescope using the 300B (MPA0843-4821 and MPA1611-4356) and 1200R (MPA1008-5202) gratings and from the AAT 2dF/AAOmega facility using the 580V and 385R volume phase holographic transmission gratings spliced together at 5700\AA\ (MPA1744-3444).}
   \label{fig:spectra}
\end{figure*}

A concerted programme of confirmatory spectroscopy was undertaken mostly during 2007 on 2-m class telescopes though six PNe were confirmed on larger telescopes during other programmes (Table \ref{tab:obs}).
In a relatively short time over 85 per cent of our refined sample of $\sim$350 PNe have been observed spectroscopically, owing in part to the higher surface brightness of many small MASH-II PNe compared to MASH. 
The majority of the spectroscopic data are confirmatory (5--20 minutes) taken with low-resolution gratings to provide sufficient wavelength coverage and to simplify subsequent reduction. 
However, some deeper spectra (40--60 minutes) have been taken of a small sample for which the faint, temperature sensitive lines were visible in the confirmatory exposure to measure chemical abundances.
A number of spectra at higher resolution were also taken for more precise radial velocity determination.
Spectral reductions were performed as described in Paper~I and the resultant spectra were used to eliminate contaminants based upon the same strict criteria adopted in Paper~I. Based on the available data we have omitted suspected symbiotic stars to be published separately.
A selection of assorted spectra and corresponding images are depicted in Figure \ref{fig:spectra} and the full catalogue is presented in Section \ref{sec:cat}.
Detailed analysis of the spectra will be presented in future papers in the series.

\begin{table*}
   \begin{flushleft}
      \begin{tabular}{llllllll}
      \hline
      Telescope & Run Date & Instrument & Wavelength & Resolution & Grating & Exposure & Observers\\
                & (dd mm yy) & & Range (\AA) & FWHM (\AA) & & Times (s) & \\
      \hline
      MSSSO 2.3-m & 19--25 05 06 & DBS& 3670--5655;5480--7515 & 4.5 & 600B/600R & 300--1800 & QAP/BM\\
      MSSSO 2.3-m & 18--22 02 07 & DBS & 3600--7400 & 6.0 & 300B & 300--1200 & QAP/AK/DJF\\
      MSSSO 2.3-m & 23--28 02 07 & DBS & 3600--7400 & 6.0 & 300B & 300--3000 & BM\\
      AAT 3.9-m & 26--27 03 07 & 2dF/AA$\Omega$ & 3700--8850 & 3.5/5.3 & 580V/385R & 2--3x1200 & QAP\\
      MSSSO 2.3-m & 06--13 05 07 & DBS & 3600--7400;5740--6750 & 6.0/1.6 & 300B/1200R & 120--3600 & BM\\
      SAAO 1.9-m & 09--15 05 07 & CCD SPEC & 3360-7520 & 7.1 & 300B & 300--2400 & QAP/AK\\
      MSSSO 2.3-m & 22--26 07 07 & DBS & 3600--7400;5740--6750 & 6.0/1.6 & 300B/1200R & 300--3600 & AK/WR\\
      VLT 8.2-m & 09--12 06 07 & FLAMES & 3620--5080;5740--8340 & 0.6 & LR1--3/LR5--7 & 120--3600 & AA/BM\\
      MSSSO 2.3-m & 19--25 08 07 & DBS & 3600--7400;5740--6750 & 6.0/1.6 & 300B/1200R & 300--1800 & AK\\
      MSSSO 2.3-m & 12--18 11 07 & DBS & 3600--7400;5740--6750 & 6.0/1.6 & 300B/1200R & 300--1800 & AK/KDP\\
      \hline
   \end{tabular}
   \textit{Notes:} Spectrograph slit generally set at 2.0--3.0 arcsec (5.0 arcsec for May 2006). 2dF fibres have an aperture of 2.1 arcsec. FLAMES mini-IFUs have 20 fibres of 0.52 arcsec aperture covering a total 2.1 $\times$ 3.1 arcsec$^2$. 
   \end{flushleft}
   \caption{Summary details of the MASH-II spectroscopic follow-up programme.}
   \label{tab:obs}
\end{table*}

\section{The MASH-II Catalogue}
\label{sec:cat}
We follow MASH closely in many aspects of the catalogue structure and dissemination of MASH-II.
Table \ref{tab:cat} shows the first five entries in the MASH-II PNe catalogue. It is available in full from VizieR and also online alongside MASH.\footnote{http://vizier.u-strasbg.fr/vizier/MASH} The catalogue format emulates the original MASH format closely with some minor changes (described below). There are over 300 PNe in the catalogue, $\sim$80 per cent of which have been spectroscopically confirmed to be bona-fide PNe, whilst a further $\sim$50 high quality candidate PNe are included which have not yet been observed. The catalogue includes four PNe with newly identified blue central star candidates, FPM0904-4023, FPM0911-4051, FPM1054-7013, and FPM1613-5633, found during the initial inspection of blocked-down H$\alpha$ images described in Paper~I, having been added after application of MASH-II visualisation techniques (Section \ref{sec:vis}) to an unpublished list of uncertain candidates. Incremental updates will take place in the same fashion as MASH, for example, to update the status of the $\sim$50 PNe without spectroscopic confirmation and to later add spectra.

\begin{table*}
   \centering
   \begin{tabular}{ccccccccclcc}
      \hline
      PN G & Name & $\alpha$ & $\delta$ & $\ell$ & $b$ & maj &	 min & CS & Morph. & Tel & Obs\\
      	  &    	& J2000  & J2000  &	 degrees 	 & degrees &	 arcsec &  arcsec & & & &  Y-M-D \\
      \hline
      216.9-05.2 & MPA0639-0554 & 06 39 58.1 & -05 54 57 & 216.9073 & -5.2373	& 10 & 10 &	- & S & MS & 2007-02-23\\
      215.7-03.9 & BMP0642-0417 & 06 42 18.4 & -04 17 49 &215.7226&-3.9861&888&560& B & Ear & MS &2007-02-20\\
      229.0-08.7 & MPA0649-1816 & 06 49 02.7 & -18 16 38 &229.0617&-8.7327&12&9& - &  Es  & MS &2007-02-23 \\
      234.9-09.7 & MPA0656-2356 & 06 56 00.0 & -23 56 49 & 234.9511 & -9.7235 & 170 & 170 & B & R & MS & 2007-11-12\\
      237.4-09.6&BMP0700-2607&07 00 51.8&-26 07 18&237.4237&-9.6662&162&50&B&A&MS&2007-08-21 \\
      \ldots & \ldots & \ldots & \ldots & \ldots & \ldots & \ldots & \ldots & \ldots & \ldots & \ldots & \ldots\\
   \end{tabular}
   \caption{The first five entries of the MASH-II catalogue that is available in full from VizieR. Entries are ordered by RA and some columns have been excluded for clarity.}
   \label{tab:cat}
\end{table*}

The following modifications and additions to column entries apply uniquely to the MASH-II catalogue. 
For a detailed description of common column entries we refer the reader to Paper~I.

%\begin{flushleft}
\emph{Object status flag: n\_PNG}\\
%\end{flushleft}
The object status flag values are the same as MASH, i.e. true (T), likely (L) and possible (P), however we include some candidates awaiting confirmatory spectroscopy as `P' where we are relatively confident of their status as PNe. 

%\begin{flushleft}
\emph{Central star: CS}\\
%\end{flushleft}
We have added this column to indicate the presence of a central star, which may take the values `B' (blue), `[WR]' (Wolf-Rayet) or `\emph{wels}' (Tylenda, Acker \& Stenholm 1993). A supplementary `?' indicates a possible blue CS or uncertain [WR] classification. 
For each PN with a `CS' entry, the equatorial and Galactic coordinates of the PN are of the central star.
Blue central star candidates were identified from a constructed B minus R difference image from SuperCOSMOS data and/or from spectra that show clear features of the central star such as a strong blue continuum or stellar lines.
Classifications of [WR] central stars are given wherever possible under the Acker \& Neiner (2003) scheme.

\subsection{General Properties}
\begin{figure*}
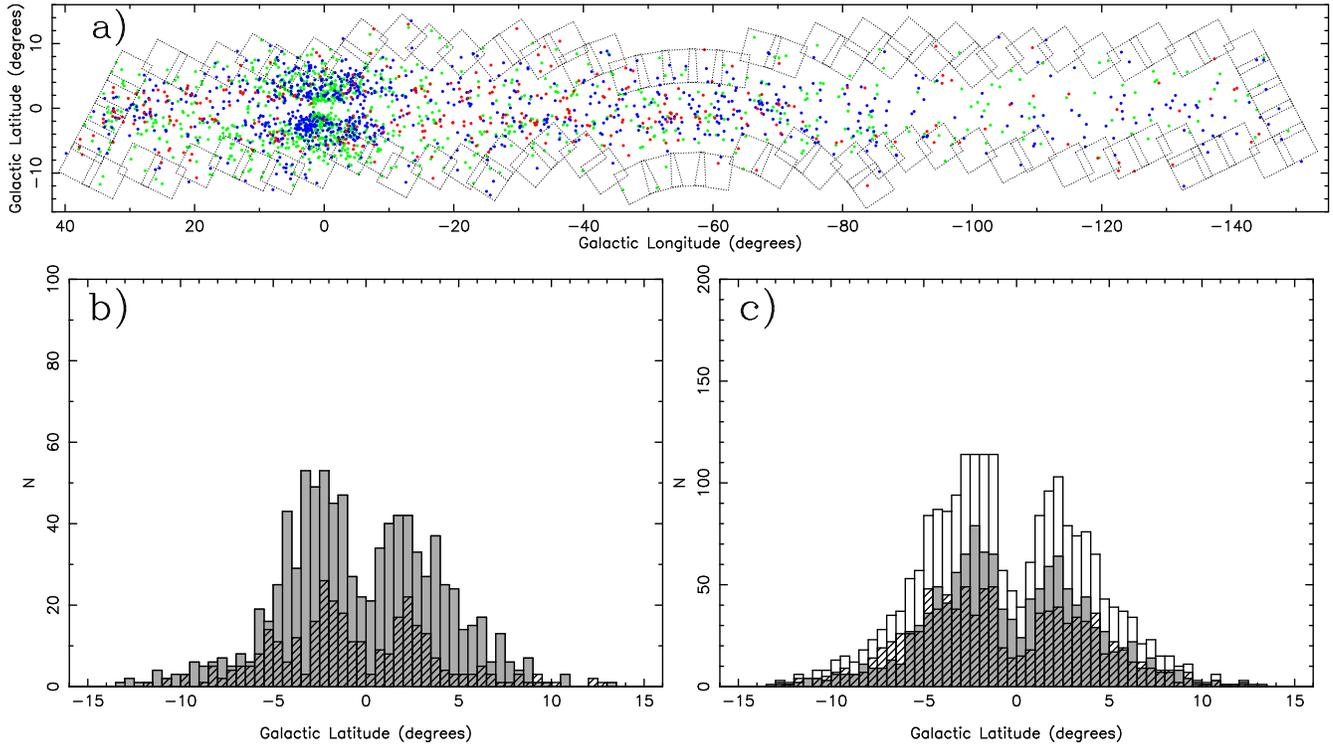

   \begin{center}
      \includegraphics[scale=0.7,angle=270]{fig6a.ps}\\
      \smallskip
      \includegraphics[scale=0.36,angle=270]{fig6b.ps}
      \includegraphics[scale=0.36,angle=270]{fig6c.ps}\\
   \end{center}
   \caption{(a) Galactic distribution of PNe covered by the SHS (dotted lines indicate peripheral fields). The PNe are sourced from non-MASH (green), MASH (blue) and MASH-II (red) catalogues. Known PNe outside the boundaries of the SHS are not shown. (b) Histogram of Galactic latitudes of MASH (solid) and MASH-II (hatched) PNe. (c) Histogram of Galactic latitudes of non-MASH (hatched), combined MASH$+$MASH-II sample (solid) and their sum representing the total population (clear).}
   \label{fig:distrib}
\end{figure*}

With MASH and MASH-II we can now present a fairer Galactic distribution of PNe (Figure \ref{fig:distrib}).
Figure \ref{fig:distrib}(a) shows the Galactic distribution of non-MASH (green), MASH (blue) and MASH-II (red) PNe covered by the SHS (dotted lines indicate peripheral SHS fields). The non-MASH sample contains an extensive compilation of $\sim$960 literature PNe from a variety of sources with positions verified by SHS images. The survey coverage is strictly complete within $-145 < \ell < 30$ and $|b| \le 6$, with substantial coverage extending to $b\sim$10 and increasingly less coverage towards higher latitudes. Some MASH PNe have been included where, although covered by the extremes of the survey, they lack online data and can be seen just outside the SHS borders in Figure \ref{fig:distrib}(a). A number of new PNe have been found independently in the $|b| \le 5$, $\ell > 30$ region overlapping with the INT Photometric H$\alpha$ Survey (Drew et al. 2005). The completeness of MASH-II is discussed in Section \ref{sec:comp}.

Additionally, Figure \ref{fig:distrib} depicts histograms comparing Galactic latitudes of both MASH catalogues (Figure \ref{fig:distrib}b), and comparing non-MASH PNe to the combined MASH$+$MASH-II sample (Figure \ref{fig:distrib}c). As noted in Paper~I, again we see the excellent sensitivity of the H$\alpha$ survey data to discover PNe at lower Galactic latitudes than previous surveys that were often focused on detecting [OIII] emission. Indeed, the new combined MASH$+$MASH-II sample is significantly larger inside $|b| \le 3$ than non-MASH PNe (Figure \ref{fig:distrib}c). The persisting deficit of PNe within $|b| \le 1$ is not unexpected for optical surveys such as the SHS and more narrow-band near-infrared surveys will be required to address the issue (see later). The higher completeness and large size of the combined MASH and MASH-II samples will have considerable impact on PN scaleheight calculations (e.g. Phillips 2001).

\begin{figure*}
   \begin{center}
      \includegraphics[scale=0.35]{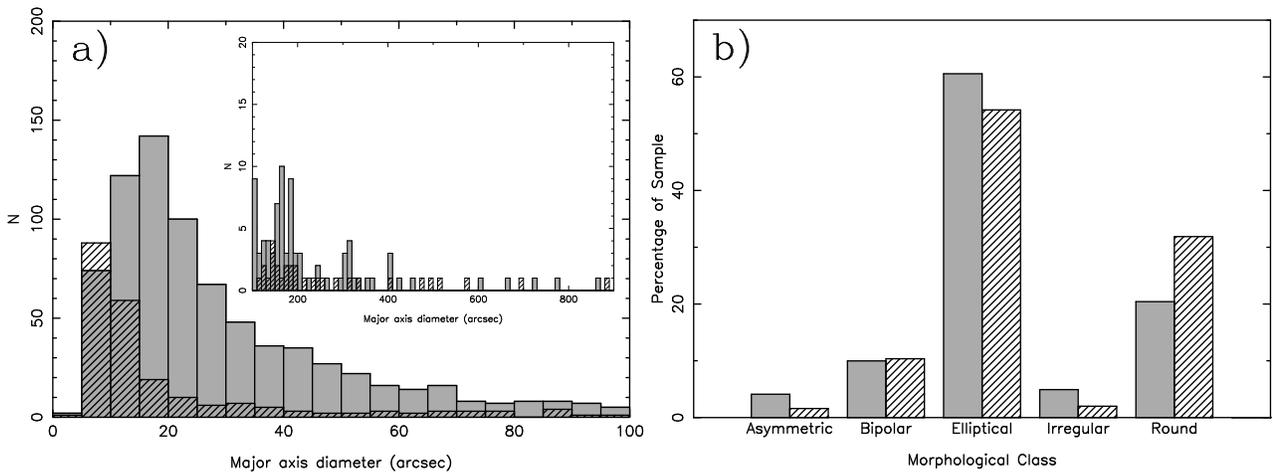}
   \end{center}
   \caption{(a) Diameters and (b) Morphologies of resolved MASH (solid) and MASH-II (hatched) PNe. The inset shows larger nebulae 100--900 arcsec in diameter. The morphological classification scheme used is described in Paper~I.}
   \label{fig:diammorph}
\end{figure*}

\begin{figure*}
   \begin{center}
      \includegraphics[scale=1.25]{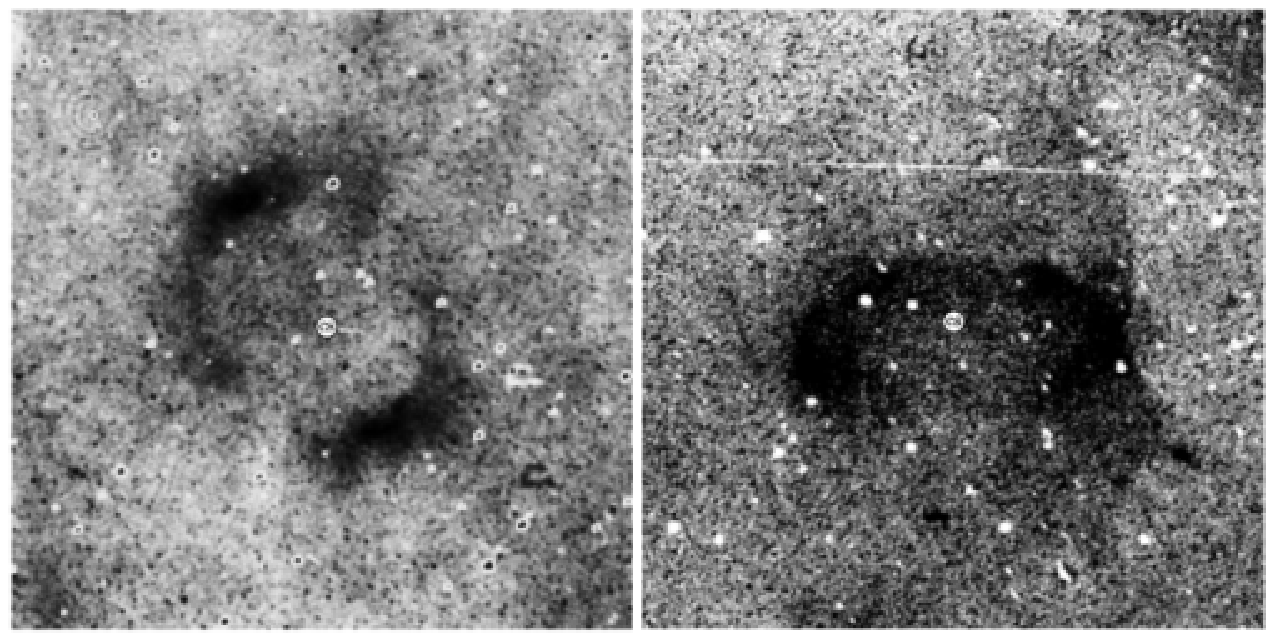}
   \end{center}
   \caption{Two 15 $\times$ 15 arcmin$^2$ quotient images of very large MASH-II PNe enhanced to show faint structure.
   BMP1808-1406 (left) exhibits elliptical geometry with opposing enhancements. BMP0733-3108 (right) shows bipolar morphology complete with faint outer lobes on both sides of an elliptical centre. Positions of blue central star candidates are marked.}
   \label{fig:largepne}
\end{figure*}

The angular diameters and morphologies of non-stellar (i.e. resolved or barely resolved) MASH and MASH-II PNe are presented in Figure \ref{fig:diammorph}. Both diameters and morphologies were determined in the same fashion as for MASH PNe. PNe smaller than $\sim$20 arcsec dominate MASH-II as a direct consequence of the IAM search (Figure \ref{fig:diammorph}a). The quotient image search has contributed many extremely low surface brightness arcminute sized PNe (see inset of Figure \ref{fig:diammorph}(a) and e.g. Figure \ref{fig:largepne}). MASH-II has added three PNe larger than $\sim$10 arcmin (one nearly 15 arcmin) to the nine existing MASH PNe in this domain (four of which exceed 15 arcmin). Such large PNe make an important contribution to the small number of local PNe, a sample critical in refining the total number, scaleheight and birth rate of PNe in the Galaxy (e.g. Frew, Parker \& Russeil 2006).

A broad, preliminary morphological comparison between resolved MASH and MASH-II PNe is shown in Figure \ref{fig:diammorph}(b). The classification scheme is that adopted in Paper~I and we include uncertain classifications (e.g. `B?'), and dual classifications (e.g. `Ep/B') for which we include only the first classification before the `/'. 
The average percentages of both MASH and MASH-II PNe are 10 per cent bipolar, 55 per cent elliptical and 28 per cent round. This is in good agreement with 13 per cent bipolar, 59 per cent elliptical and 28 per cent round seen by Manchado et al. (2000) in a large sample of Northern Galactic Plane PNe for $|b| \le 4$. We independently arrive at a 10 per cent bipolar fraction from both MASH and MASH-II PNe. The value may be closer to the 12.5--15 per cent reported in Paper~I after more careful treatment of dual classifications and reclassification of asymmetric and irregular PNe into the other classes.

As we have not performed any PSF fitting of MASH or MASH-II PNe prior to morphological classification, the number of round and elliptical PNe may be different to the preliminary analysis given here as there can be some field rotation during 3 hour SHS exposures. These numbers for MASH-II PNe may also be affected by some of our cuts that favoured round PNe, however the number of PNe found from these cuts were small and these cuts may also be dependent on the ellipticity of deblended profiles which can be skewed by nearby stars. Figure \ref{fig:diammorph} also suggests that MASH-II PNe are smaller and rounder than MASH PNe, on average, and are therefore more distant with larger $|z|$ distances since round PNe are generally found further from the Galactic Plane (Manchado et al. 2000;Phillips 2001). A more detailed morphological analysis will be presented in future papers.

\section{Completeness of MASH-II}
\label{sec:comp}
We are confident MASH-II has found the majority of PNe undiscovered in the SHS as described in Paper~I.
Our discovery techniques were designed to find these PNe efficiently without resorting to downloading full-resolution data for each of the 233 SHS fields. Such efficiency does mean that a small number of PNe, may still be found in the SHS, but these will either lie outside our search criteria or be generally difficult to find requiring tedious analysis of large amounts of data. In the following we discuss limitations of our techniques and suggest some future directions.

The IAM photometry search has been efficient in finding generally small, bright, star-like or resolved PNe. However, the photometry is ultimately limited by the deblending algorithm used and its associated parameters. Our chosen parameters minimised our losses as evident in discoveries of PNe near bright stars (Figure \ref{fig:catquality}) and a small number of extremely faint PNe based on deblending of their bright rims (e.g. MPA1834-2222). We have discussed that the separate deblending of H$\alpha$ and SR wavebands limits the utility of H$\alpha-$SR cuts which can be undefined in some cases (Figure \ref{fig:nosr}) and we have begun to address this with our onerous fourth cut (Table \ref{tab:cuts}). Yields from this cut have been small and consequently it has only been applied to 78 per cent of SHS fields. A small number of faint or highly elliptical PNe may also be found outside our H$\alpha-$SR cuts, however it will be difficult to discern any true faint PNe from scanned film artefacts such as dust and emulsion defects (see Section 8 of Parker et al. 2005 for full details). We are currently considering further cuts to target these `missed' PNe, however for the reasons outlined above, we suspect tedious inspection of full-resolution and blocked-down pixel data close to the plane will prove to be more useful. 

Our use of blocked-down quotient images of fields has enabled many exceedingly faint, large PNe to be found. Though there are some important limitations. 
Of main concern is the relatively coarse resolution of 11 arcsec per pixel. This resolution reduces sensitivity to PNe less than two arcminutes across, although if there is little background emission they may be visible as small groups of pixels. Early on in the quotient image search some groups of pixels were investigated (many were also artefacts), but this was later discontinued in favour of the more efficient IAM search. Therefore, if some of these detections were too faint to be detected by the IAM search, they may have been missed. One such example might be MPA1602-5543 (found serendipitously at full resolution) that is just visible in the quotient image but too faint for IAM photometry to deblend it. We are currently addressing this issue by further searches, however higher resolution blocked-down images (e.g. 5 arcsec per pixel) would help to reduce the background contribution and increase sensitivity to small, faint PNe.

Furthermore, only one of us (JB) has gone through all the blocked-down quotient images systematically. 
This is not an ideal situation as the wealth of (often complex and overlapping) detail in these large images, combined with the often subjective nature of identifying the diverse morphologies of PNe, can lead to some objects being missed unless the work is carefully repeated by others. One of us (BM) has independently viewed a sample of $\sim$30 fields. Only very few additional PNe were recovered from this search and often in unusual locations, e.g. along a field edge (MPA0656-2356). The search by JB also produced many candidates of atypical or irregular morphology, some of which are included in MASH-II, however most turned out to be HII regions. More detailed study of these candidates may produce a small number of true PNe, though we note our visualisation technique helped in many cases to identify blue central star candidates. We did not flat-field the blocked-down H$\alpha$ and SR images prior to division (see Parker et al. 2005) as it seemed to degrade our sensitivity to the periphery of each field. Analysis of positions of identified PNe showed no bias towards any field region that may have resulted from non flat-fielded data. The film media of field HA273 was too heavily scratched to identify any new PNe. 

The few PNe remaining to be found within the SHS will be relatively difficult to find. Perhaps the most promising means to finding more (highly reddened) PNe lies with narrow-band imaging surveys in the [SIII] line $\lambda$9532\AA\ (e.g. Kistiakowsky \& Helfand 1995, Jacoby \& Van de Steene 2004). A more extensive [SIII] survey covering the whole Galactic Bulge and the Galactic Plane ($|b| \le 2$) is needed. Some of these PNe are visible just in the noise level of the SHS. As for PNe outside the SHS coverage, there likely exists a considerable number of faint PNe 1--5 arcmin across out to $b\sim$20 degrees. The SHS has enabled discovery of some of these (e.g. MPA0656-2356 at $b\sim-$10), however they are too small for the coarse 48 arcsec pixels of SHASSA (Gaustad et al. 2001; Frew, Madsen \& Parker 2006). Some have been found with broad-band sky surveys (e.g. Kronberger et al. 2006; Jacoby et al. 2007), however many are of too low surface brightness to be sufficiently visible to enable routine discovery using broad-band surveys. An extension of the SHS to higher latitudes with new survey telescopes such as SkyMapper (Keller et al. 2007) will enable such discoveries to be conducted efficiently in the future.

\section{Conclusions}
\label{sec:conc}
The MASH-II catalogue of over 300 true, likely and possible PNe has been presented. 
We provide accurate positions, angular diameters and morphological classifications for all catalogue entries. Over 80 per cent of entries have been observed spectroscopically. The new PNe have been discovered by application of semi-automated discovery techniques applied to the online digital data of all 233 SHS fields for the first time. We used IAM photometry to target high surface brightness, small PNe and blocked-down quotient images of SHS fields to target low surface brightness, large PNe. Candidates were visualised using multi-wavelength data, making particular use of UKST SuperCOSMOS B$_\mathrm{J}$ images at the earliest stages to increase our detection capability. This enabled many PNe to be found that may have escaped detection with simpler `two-colour' discovery techniques. MASH-II will strengthen the many science programs already underway with MASH (see Paper~I for details).

\section*{Acknowledgments}
BM, AK and DJF acknowledge Macquarie University for the provision of PhD scholarships.
BM further acknowledges Observatoire de Strasbourg and PICS for support.
JB acknowledges the Anglo-Australian Observatory for a vacation scholarship to conduct the large nebulae search. We thank Francois Ochsenbein for assistance in establishing the online component of MASH-II at the Centre de Donn\'ees astronomiques de Strasbourg (CDS). We thank the referee for useful comments. We thank the South African Astronomical Observatory, Australian National University and European Southern Observatory telescope time allocation committees. 
We thank the ANSTO access to major research facilities scheme for supporting QAP and AK to conduct SAAO observations, and ESO/Observatoire de Strasbourg for travel support for the VLT run. We thank Warren Reid (WR) and Kyle DePew (KDP) for assisting AK with the July 2007 and November 2007 2.3-m observing runs respectively.
This research has made use of the SAOImage DS9, developed by Smithsonian Astrophysical Observatory, the SIMBAD database operated at CDS, Strasbourg, France, and makes use of data products from the 2MASS survey, which is a joint project of the University of Massachusetts and the Infrared Processing and Analysis Center/California Institute of Technology, funded by the National Aeronautics and Space Administration and the National Science Foundation.

\end{document}